\documentstyle[aps,floats,epsfig]{revtex}
\begin{document}

\newcommand{\bce}{\begin{center}}
\newcommand{\ece}{\end{center}}
\newcommand{\boldF}{\mbox{\boldmath $F$}}
\newcommand{\boldV}{\mbox{\boldmath $V$}}
\newcommand{\boldpi}{\mbox{\boldmath $\pi$}}
\def\lsim{\mathrel{\rlap{\lower4pt\hbox{\hskip1pt$\sim$}}
    \raise1pt\hbox{$<$}}}       
\def\gsim{\mathrel{\rlap{\lower4pt\hbox{\hskip1pt$\sim$}}
    \raise1pt\hbox{$>$}}}

\title{Photon and dilepton emission rates from high density quark matter}

\author{Prashanth Jaikumar, Ralf Rapp and Ismail Zahed} 

\address{Department of Physics and Astronomy, SUNY at Stony Brook, 
New York 11794-3800}

\date{\today} \maketitle

\begin{abstract}
We compute the rates of real and virtual photon (dilepton) emission
from dense QCD matter in the color-flavor locked (CFL) phase,
focusing on results at moderate densities (3-5 times the nuclear 
saturation density) and temperatures $T\simeq80$~MeV. 
We pursue two approaches to evaluate the electromagnetic (e.m.) 
response of the CFL ground state:
(i) a direct evaluation of the photon self energy using quark particle/-hole
degrees of freedom, and 
(ii) a Hidden Local Symmetry (HLS) framework based on generalized mesonic
excitations where the $\rho$ meson is introduced as a gauge boson of a 
local SU(3) color-flavor group. The $\rho$ coupling to generalized two-pion 
states induces a finite width and allows to address the issue of vector 
meson dominance (VMD) in the CFL phase.    
We compare the calculated emissivities (dilepton rates) 
to those arising from standard hadronic approaches including
in-medium effects.   
For rather large superconducting gaps (several tens of MeV at moderate
densities), 
as suggested by both perturbative and nonperturbative estimates,  
the dilepton rates from CFL quark matter turn out to be very similar to 
those obtained in hadronic many-body calculations, especially for
invariant masses above $M\simeq0.3$~GeV. 
A similar observation holds for (real) photon production.
\end{abstract}
\section{Introduction}
\label{sec_intro}
Recent extensive studies of QCD at finite quark chemical potential 
($\mu_q$) have established that the ground state at large densities 
and zero temperature is a color superconductor with large pairing gaps
on the order of 100~MeV (see, e.g, ref.~\cite{Krishna} and references
therein). Specifically, for 3 flavors of massless quarks,
general arguments based on free energy minimization and maximal
symmetry favor the formation of a condensate that effectively locks
rotations in color and flavor space~\cite{ARW99,RSSV00}. Among the 
intriguing implications of this  color-flavor locking (CFL) are  
the modifications of photon propagation in such matter, e.g.,   
true electromagnetism being a combined rotation of the standard 
$U(1)_{Q}^{em}$ and $U(1)_Y$ color hypercharge. The color-flavor locked 
medium is thus transparent to this rotated photon~\cite{Kris}. More 
recent studies have addressed the question of reflection and refraction at 
the interface of normal (pairing-free) and CFL media~\cite{Manu}. 
Given a finite (small) temperature, a natural question would be to ask for 
the blackbody radiation of the system, i.e., 
how much a block of CFL matter shines via emission of real photons or
dileptons (virtual photons). For photons, the answer is pertinent to
their propagation in the cores of neutron stars, which
may be sufficiently dense for formation of color superconducting quark
matter. A study of the temperature dependence of the photon
emissivities and mean free paths in the CFL phase can determine how
transparent this phase is at various temperatures.

\vskip 0.2cm 
Apart from consequences for specific physical systems like neutron stars, 
one ought to address the issue of photon and dilepton emission at varying  
temperature and baryon density as a general means to investigate
different regions of the QCD phase diagram, especially in colder and
denser regions. The latter may be accessible to the HADES experiment 
at GSI~\cite{Hades}.
Small final state interactions make photon and dilepton probes  
important long lived signatures of the state of strongly interacting matter. 
Corresponding production rates have been extensively studied for both hot 
hadronic matter incorporating medium 
effects~\cite{GK91,CS93,HFN93,GL94,CRW96,SYZ1,SYZ2,KKW97} 
(see ref.~\cite{RW00} for a recent review) as well as for the Quark-Gluon 
Plasma using perturbative~\cite{MT85,CFR87,AA89,BPY90,PT00} 
and nonperturbative~\cite{Z90,LWZH98,ST98} approaches. 
It has been pointed out that perturbative $q\bar{q}$ annihilation rates in a QGP 
at comparable temperatures and densities are suggestive of a 
'quark-hadron duality', i.e., the dilepton signature for an interacting 
confined phase with broken chiral symmetry resembles that of a QGP with chiral symmetry 
restored. This has been found for both the low-mass ($\rho$-/$\omega$-) region~\cite{RW99}
and the intermediate-mass region~\cite{LR98,LYZ98} (between the $\phi$ and $J/\Psi$). 
Could such a dual scenario hold for the dense 
phase of condensed diquarks at finite temperature as well? 
In this paper, we will address this question by making perturbative estimates
of the photon and dilepton emissivities from dense superconducting matter 
with 3 flavors of massless quarks (the CFL phase). While such perturbative 
calculations are strictly valid only at parametrically large chemical 
potential, it is intriguing to extrapolate down to densities of a few times 
that of normal nuclear matter ($n_0=0.16$~fm$^{-3}$). 
In this way, our approach attacks the finite density problem from the side 
of large baryon density and serves to complement studies of emission rates 
from the low density (hadronic) side, extended to include medium effects. 
Significantly, our calculation at finite density
allows for finite temperature effects that are possibly as large as
the gaps in the CFL phase, conjectured to be several tens of MeV at
$\mu_q\simeq$~300-500 MeV.

\vskip 0.2cm 
Another issue we have sought to address in our analysis of the dilepton 
emission rates is the viability of vector meson dominance (VMD) at high
density. Along the lines of ref.~\cite{Ho99,RSWZ00} we have included 
excitations describing the rho meson ($\rho$) in the CFL phase via an effective
theory based on hidden local symmetry. Based on weak coupling analyses to 
leading logarithm accuracy, effective Lagrangian approaches to
QCD in the CFL phase have proven to be useful with
applications to color-flavor anomalies, hidden local
symmetry and meson properties~\cite{Ho99,Casal,Son}. Here, we use the 
effective Lagrangian to obtain the coupling of the $\rho$ meson to the 
rotated photon in the CFL phase and model its width in the dense phase 
via a Breit-Wigner resonance through its coupling to two generalized pions. 
In this way, we recover VMD and show that the $\rho$ meson makes an 
important contribution to the dilepton emission rate, much like in 
low density hadronic frameworks.  

\vskip 0.2cm 
The organization of this paper is as follows. In section~\ref{sec_cc1}, we 
compute the photon polarization tensor from the quark loop in the CFL phase
(independently confirming the result first obtained in ref.~\cite{LM01}),
and present numerical estimates for corresponding dilepton
and photon rates at finite temperature and for baryon densities down to
$n_B=$~(3-5)~$n_0$. In section~\ref{sec_cc2}, we derive the
effective Lagrangian for the CFL phase which incorporates hidden local
symmetry and show how the generalized $\rho$ meson arises as a dynamical
gauge boson of this symmetry. In particular, we address the fate of 
vector meson dominance in this model, thereby determining the coupling of 
the generalized vector mode to the photon. Again, we present the associated
contribution to the dilepton production rate. 
In section~\ref{sec_comp}  we confront the combined emissivities from CFL 
matter with prior estimates from in-medium hadronic as well as perturbative 
QGP calculations. 
We summarize and conclude in section~\ref{sec_concl}. 
 
\section{CFL Phase and E.M. Current Correlator I: Weak Coupling} 
\label{sec_cc1}
\subsection{One-Loop Photon Self Energy}
The photon polarization tensor in color-flavor locked quark matter 
to one-loop order was first derived by Litim and Manuel~\cite{LM01}. 
For the sake of completeness, and as an independent confirmation
of their result, we here provide a calculation using a different quark
basis than in ref.~\cite{LM01}. In what follows we adopt the notation of ref.~\cite{Rischke}. 
We assume all quarks to be massless. The self energy receives contributions 
from the diagonal (dressed quasiparticles) and off diagonal (diquark) 
components of the Nambu-Gorkov propagator which reads~\cite{pisarski}
\begin{equation}
S(K) = \biggl(\begin{array}{cr} G^{+}(K) & \Xi^{-}(K) \\ \Xi^{+}(K) &
G^{-}(K)
\end{array}\biggr) \ ,  
\label{invprop}   
\end{equation}
where $(G^{\pm})^{-1}\equiv G_{0}^{\pm}-\Sigma^{\pm}$, \ 
$\Sigma^{\pm}\equiv \Phi^{\mp}G_{0}^{\mp}\Phi^{\pm}$, \ 
with
$\Phi^{+}\sim\langle\psi_C\bar{\psi}\rangle$,
$\Phi^{-}\sim\langle\psi\bar{\psi}_{C}\rangle$ and
$(G_{0}^{\pm})^{-1}(K)=K{\hskip-2.7mm}/\pm\mu_q\gamma_{0}$. $\psi_C$
denotes the charge conjugate partner of $\psi$. The off-diagonal
components of the matrix in eq.(\ref{invprop}) are $\Xi^{\pm}\equiv -G_0^{\mp}\Phi^{\pm}G^{\pm}$.
With the convenient choice of $SU(3)$ generators
$T^8=\lambda_8/2=(\sqrt{3}/2)Q, \,(Q={\rm
diag}\{2/3,-1/3,-1/3\})$ and $T^3={\rm diag}\{0,1/2,-1/2\}$, the
mixing of the photon with the eighth gluon can be described by the linear
combinations~\cite{LM01}
\begin{eqnarray}
\tilde{A}_{\mu}&=&A_{\mu}{\rm cos}\,\theta_{CFL}- G_{\mu}^{8}{\rm
sin}\, \theta_{CFL}\\ \tilde{G}_{\mu}^{8}&=&A_{\mu}{\rm sin}\,
\theta_{CFL}+G_{\mu}^{8}{\rm cos}\, \theta_{CFL} \quad,
\end{eqnarray}
where ${\rm tan}\,\theta_{CFL}= 2e/\sqrt{3}g$, and $e$, $g$ denote the
gauge couplings of $A_\mu$, $G_\mu$ respectively. The CFL quarks couple to 
the in-medium photon $\tilde{A}_{\mu}$ with strength $\tilde{e}=e\,{\rm cos}\, \theta_{CFL}$
via the vertex
\begin{equation}
(\Gamma^{\mu})^{ij}_{ab}=(\tilde{Q})^{ij}_{ab}\gamma^{\mu}=
\frac{1}{\sqrt{3}} \biggl[(\lambda_{8})^{ij}\delta_{ab}
-\delta^{ij}(\lambda_{8})_{ab}\biggr]
\gamma^{\mu} \quad,
\label{tildep}
\end{equation}
since $\tilde{Q}=Q\otimes 1+1\otimes Y$ where $Y$ represents the hypercharge
operator. In eq.~(\ref{tildep}), $\{i,j\}$ and $\{a,b\}$ denote flavor- 
and color-indices, respectively.  The self-energy then takes the form 
\begin{equation}
\Pi^{\mu\nu}(K)=\frac{\tilde{e}^2}{2}{\rm Tr}_{k,s,c,f,NG}
\biggl[\Gamma^{\mu}S(k+\frac{K}{2})\Gamma^{\nu}S(k-\frac{K}{2})\biggr] \ ,
\end{equation}
where the traces are to be carried out over internal 4-momentum ($k$),
spin ($s$), color ($c$), flavor ($f$) and Nambu-Gorkov ($NG$) indices. 
The $NG$ trace is easily performed in view of eq.~(\ref{invprop}), 
whereas the color-flavor locked structure of the
condensate is conveniently unraveled using the projectors~\cite{Shovy}, 
\begin{equation}
C^{{1}^{ij}}_{ab}\equiv \frac{1}{3}\delta^i_a\delta^j_b,\quad
C^{{2}^{ij}}_{ab}\equiv \frac{1}{2}(\delta_{ab}\delta^{ij}
-\delta^j_a\delta^i_b),\quad C^{{3}^{ij}}_{ab}\equiv
\frac{1}{2}(\delta_{ab}\delta^{ij}
+\delta^j_a\delta^i_b)-C^{{1}^{ij}}_{ab} \ .
\end{equation}
These projectors may be identified with the singlet ($P_1$) and
octet ($P_8$) ones~\cite{Zar} according to
\begin{equation}
P_1=C^1,\quad P_8=C^2+C^3 \ .
\end{equation}
Using the following trace relations, 
\begin{eqnarray}
{\rm Tr}_{c,f}\biggl[\tilde{Q}P_1\tilde{Q}P_1\biggr]&=&0={\rm Tr}_{c,f}\biggl[\tilde{Q}P_1\tilde{Q}P_8\biggr]=
{\rm Tr}_{c,f}\biggl[\tilde{Q}(C^2-C^3)\tilde{Q}^{T}P_1\biggr]\\ \nonumber 
{\rm Tr}_{c,f}\biggl[\tilde{Q}P_8\tilde{Q}P_8\biggr]&=&4=-{\rm Tr}_{c,f}\biggl[\tilde{Q}(C^2-C^3)\tilde{Q}^{T}(C^2-C^3)\biggr] \ , 
\end{eqnarray}
the self energy of the CFL photon can be expressed in compact form as
\begin{equation}
\Pi^{\mu\nu}(K)=2\tilde{e}^{2}\biggl[G_{88}^{\mu\nu}(K)+\Xi_{88}^{\mu\nu}(K)
\biggr]  \ ,
\end{equation}
with
\begin{eqnarray}
G_{88}^{\mu\nu}(K)&=&\sum_{e=\pm} {\rm Tr}_{k,s}
\biggl(\gamma^{\mu}G^{e}(k+\frac{K}{2})\gamma^{\nu}G^{e}(k-\frac{K}{2})\biggr)
\\ 
\Xi_{88}^{\mu\nu}(K)&=&\sum_{e=\pm} {\rm Tr}_{k,s} \biggl(
\gamma^{\mu}\Xi^{e}(k+\frac{K}{2})\gamma^{\nu}\Xi^{-e}(k-\frac{K}{2}) 
\biggr) \ . 
\end{eqnarray}
This result for $\Pi^{\mu\nu}$ explicitly agrees with the one quoted in 
ref.~\cite{LM01}, cf.~eq.(27) therein. The subscript $88$ indicates
that only the octet projection contributes. The sum over $e=\pm$ encompasses 
the contributions of particles (+) and antiparticles (-). As the 
antigap dependent pieces arise only at order 
${\cal O}(\mu_q^{-2})$~\cite{RSWZ00}, 
we neglect pairing effects for antiquarks in our calculation of
emission rates. Moreover, the antigap is 
gauge dependent even for forward scattering of antiquarks~\cite{schaf}. We
keep, however, the contribution from free antiquarks. This, in particular,
has the advantage of naturally recovering the dilepton rate from the 
perturbative $q\bar{q}$ annihilation process for vanishing quark gaps, 
which is our baseline for assessing the effects of pairing. The Dirac spin 
traces may be performed explicitly using equations (45) and (95) of 
ref.~\cite{Rischke}. The 4-momentum trace over $k$ involves a 3-momentum
integration as well as a summation over Matsubara frequencies in the 
energy component $k_0$ within the imaginary time formalism to incorporate
finite temperature effects. This allows to obtain explicit expressions  
for the various components of $\Pi^{\mu\nu}(K)$ (see ref.~\cite{LM01} for 
details), which in the following section will be employed to compute 
photon and dilepton emission rates.

\subsection{Photon and Dilepton Rates}
\begin{figure}[!h]
\vspace{0.5cm}
\bce
\epsfig{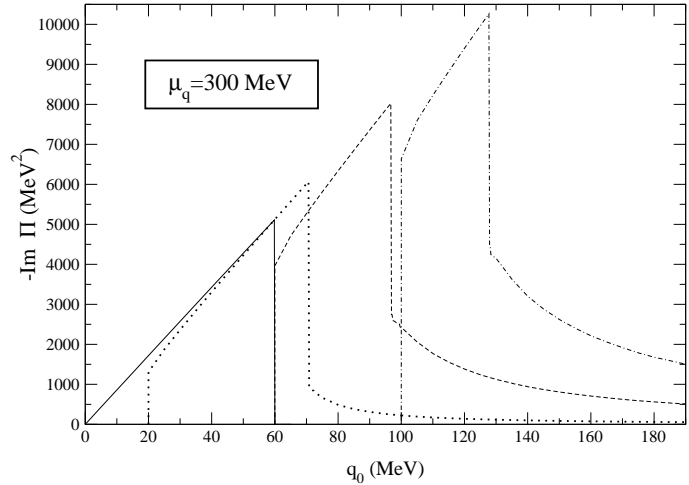}
\ece
\vspace{0.3cm}
\caption{{\rm Im}$\Pi$ at temperature $T=0$ MeV, three momentum $q=60$ MeV and
for $\Delta=0$ MeV (solid), $\Delta=10$ MeV (dotted), $\Delta=30$ MeV (dashed)
and $\Delta=50$ MeV (dash-dotted). In each case, the response has a
threshold at $q_0=2\Delta$.}
\label{figps1}
\end{figure}

\begin{figure}[h]
\vspace{0.8cm}
\bce
\epsfig{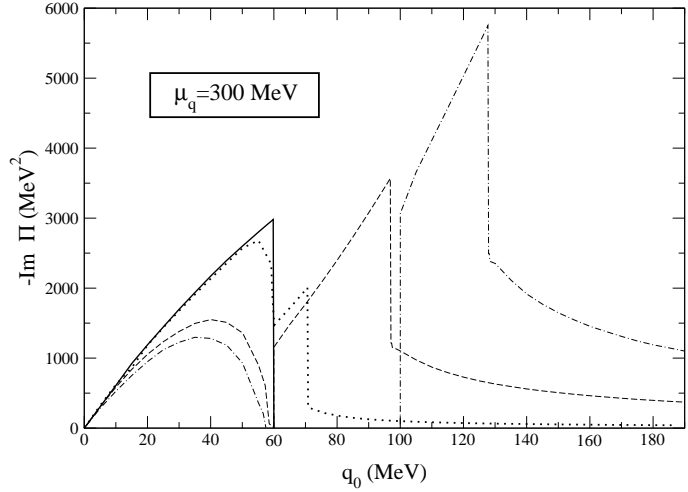}
\ece
\vspace{0.3cm}
\caption{{\rm Im}$\Pi$ at finite temperature $T=50$~MeV, three momentum 
$q=60$~MeV and for $\Delta=0$ MeV (solid), $\Delta=10$~MeV (dotted), 
$\Delta=30$~MeV (dashed) and $\Delta=50$~MeV (dash-dotted). The relocation of 
strength from the spacelike to the timelike region is evident. This signals 
the possibility of dilepton emission.}
\label{figps2}
\end{figure}

The medium modification of the photon field\footnote{We refer here to the 
(massless) rotated photon in the CFL phase~\cite{ABR00} which plays the 
role of true electromagnetism.} in the CFL phase is contained in the 
components of the polarization tensor, which in turn can be related to e.m. 
emissivities~\cite{MT85,GK91}.
The thermal emission rate of real photons reads 
\begin{equation}
q_0\frac{dR_\gamma}{d^3q}=-\frac{1}{(2\pi)^3} \ 
{\rm Im}{\Pi^{\rm (ret)}}^{\mu}_{\mu}(q_0=q) \ f^B(q_0;T) \ ,
\label{Rphot}
\end{equation}
while that for dileptons (virtual photons) is given by
\begin{equation}
\frac{dR_{ee}}{d^4q}=-\frac{\alpha}{12\pi^4 M^2} \
{\rm Im}{\Pi^{\rm (ret)}}^{\mu}_{\mu}(q_0,q) \ f^B(q_0;T) \ ,
\label{Rdilep}
\end{equation}
where $f^B(q_0;T)=1/[{\rm exp}(q_0/T)-1]$ is the thermal Bose factor,
$\alpha=e^2/4\pi$ with $e$ the usual electromagnetic
charge, and $M^2=q_0^2-q^2$ the invariant mass of the lepton pair. 
The factor of $\alpha$ implies that we choose the
projection along the ordinary photon which couples to the dileptons.
 The imaginary part of the retarded photon self energy, 
Im${\Pi^{\rm (ret)}}^{\mu}_{\mu}={\rm Im\Pi}^L+2{\rm Im\Pi}^T$, 
is the weighted sum of its transverse and longitudnal projections 
(at the photon point, i.e., for  vanishing invariant mass $M^2=q_0^2-q^2$,
$\Pi^L$ vanishes). It is obtained from $\Pi^{\mu\nu}(K)$ by  
performing the appropriate analytic continuation to arbitrary complex  
values of the external frequency $i K^{0} \equiv q_0$. The one-loop
result for positive $q_0>0$ then reads
\begin{eqnarray}
&{\rm Im}&{\Pi^{\rm (ret)}}^{\mu}_{\mu}=
\frac{\tilde{e}^2}{2\pi^2} \sum_{e_1,e_2=\pm} \int d^3k
\nonumber \\
&\times& \left\{ \frac{\Delta^2}{2}
\biggl(\frac{{\hat{\bf k}_1}\cdot{\hat{\bf k}_2}-1}
{\epsilon_1\epsilon_2}\biggr)
\biggl[\delta(q_0-\epsilon_1-\epsilon_2)\,(1-N_1-N_2)+
[\delta(q_0-\epsilon_1+\epsilon_2)-\delta(q_0+\epsilon_1-\epsilon_2)]
(N_1-N_2)\biggr]  \right.  
\nonumber \\
&& \ - \left.
\biggl[\delta(q_0-\epsilon_1-\epsilon_2)\,(1-N_1-N_2)
\biggl(1-\frac{\xi_1\xi_2}{\epsilon_1\epsilon_2}\biggr)+
\left[\delta(q_0-\epsilon_1+\epsilon_2)-\delta(q_0+\epsilon_1-\epsilon_2)\right]
(N_1-N_2)\biggl(1+\frac{\xi_1\xi_2}{\epsilon_1\epsilon_2}\biggr)\biggr] 
\right\} \ , 
\label{response}
\end{eqnarray}
where the quasiparticle energies are denoted by
$\epsilon_i=\sqrt{\xi_i^2+\Delta^2}$, with $\xi_i=e_i k_i-\mu_q$ and
$k_1=|{\vec k}|$, $k_2=|\vec q- \vec k|$ the quasiparticle momenta. 
$N_i=f^F(\epsilon_i;T)=1/[\exp(\epsilon_i/T)+1]$ are the usual Fermi
distribution functions which become $\Theta$-functions in the $T$=0-limit.
The summation over all combinations of $e_1=\pm$ and $e_2=\pm$ generates
a total of 4 contributions embraced in curly brackets.  However, since we 
approximate antiparticles as free (ungapped), the $\Delta^2$ term is zero
except for the $e_1=e_2=+$. The physical interpretation of the various 
terms  is encoded in the delta functions which determine the on-shell  
kinematics of the process (in eq.~(\ref{response}), we have omitted terms 
involving delta functions of the form $\delta(q_0+\epsilon_1+\epsilon_2)$, as
they do not contribute to the production of on-shell photons or lepton pairs).
Let us focus on the contribution of the particles to ${\rm Im} \Pi$.  
At zero temperature, Pauli Blocking enforces the  distribution functions
to vanish identically, i.e., $N_1=N_2=0$. Thus the only contribution
to the imaginary part arises from the
$\delta(q_0-\epsilon_1-\epsilon_2)$-terms.
For vanishing gap (free quark gas), energy-momentum conservation
in the particle-hole excitation implies $M^2=-2 k_1k_2(1+\cos\theta)$ 
($\theta=\angle(\vec k_1,\vec k_2)$), i.e., the response is strictly 
spacelike which corresponds to standard Landau damping (this obviously 
remains true for $T>0$). For a finite gap, the response sets in at $q_0>2\Delta$,
the break-up of a Cooper pair acting as the threshold for excitations, as 
expected. More importantly, the response extends into the timelike region as 
well, which again follows from energy-momentum conservation: choosing, e.g., 
$\cos\theta=-1$ (which is the optimal configuration for a timelike response), 
one finds $M^2=2(\Delta^2+\epsilon_1\epsilon_2-\xi_1\xi_2)$
to be positive, thus allowing for photon and dilepton emission. 
These features are confirmed in Fig.~\ref{figps1}, where 
Im~$\Pi(q_0,q,\Delta;T=0)$ is plotted versus energy at fixed
3-momentum $q=60$~MeV and for various values of the gap parameter.\\
At finite temperature, $N_1,N_2\neq 0$ and the delta functions
with energy arguments $q_0\mp\epsilon_1\pm\epsilon_2$ also contribute. 
The resulting response function as shown in Fig.~\ref{figps2} indicates  
a significant relocation of strength from the spacelike to the timelike 
region. 
Such a feature was also found  
in ref.~\cite{greg}, where neutrino scattering in the color superconductor 
has been studied. The delta functions with opposing signs on the 
quasiparticle energies are restricted to spacelike $M^2$, and are relevant
for scattering of  CFL photons off quarks (Landau damping in the CFL phase). 
In our context of photon and dilepton emission, where $M^2\ge0$, 
these terms are of no direct concern. 
However, we may note that the shifting of response strength from the
spacelike to the timelike region implies the suppression of scattering as compared to the 
free Fermi gas~\cite{greg}. 
 
\begin{figure}[h]
\vspace{0.8cm}
\bce
\epsfig{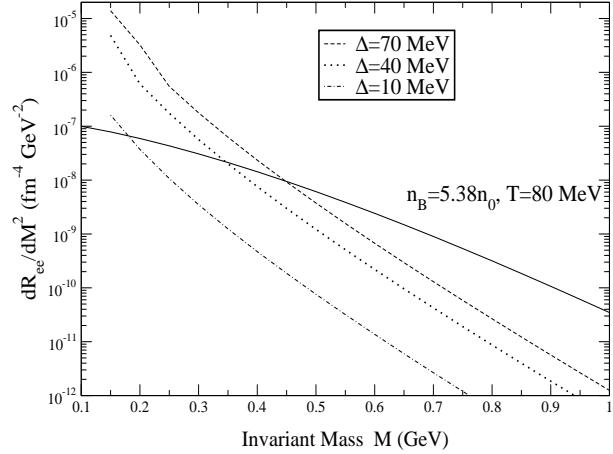}
\ece
\vspace{0.3cm}
\caption{Dilepton rates at finite gap and finite temperature (with
particles only). The rate vanishes as the gap goes to zero. The
solid line is the ${\cal O}(\alpha_S^0)$ estimate of the free
$q\bar{q}$ annihilation rate at the same temperature and density.}
\label{figps3}
\end{figure}

\begin{figure}[ht]
\vskip 1cm
\bce
\epsfig{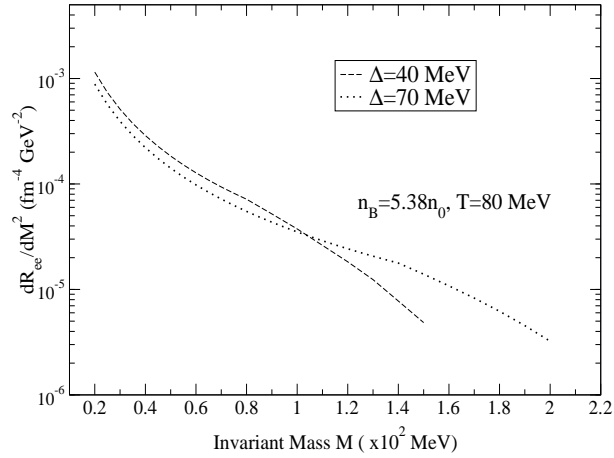}
\ece
\caption{Dilepton emission rate in the very low mass region. A smooth
turning over seems to occur at $M=2\Delta$.}
\label{figps4}
\end{figure}

To display the associated dilepton emission rates, 
it is common to focus on the three-momentum integrated quantity which produces
the invariant-mass spectrum 
\begin{equation}
\frac{dR_{ee}}{dM^2}=\int\frac{d^3q}{2q_0}\frac{dR_{ee}}{d^4q} \ .\label{imasspec}
\end{equation}
The pertinent numerical results for the CFL phase based on the one-loop
photon self energy, eq.~(\ref{response}), are displayed in Fig.~\ref{figps3}, 
where only the particle contributions are retained to highlight the impact 
of Cooper pairing on the emission rate. 
The curves have been evaluated for a quark chemical potential of 
$\mu_q=350$~MeV (corresponding to a baryon density $n_B\simeq5.4n_0$
including massless strange quarks)
assuming a temperature dependence of the gap as inferred from BCS theory, 
\begin{equation}
\Delta=\Delta_0\sqrt{1-\biggl(\frac{T}{T_c}\biggr)^2} \ .  
\label{tdep}
\end{equation}
The rather large variation in the considered gap values (10-70~MeV) is
motivated by  the sharp curvature of the above relation near
$T_c$. For definiteness, using $T_c\simeq 0.56 \Delta_0$ with  
$\Delta_0 \simeq 150$~MeV at
$\mu_q\simeq 350$~MeV (as suggested by both perturbative~\cite{schafer}
and nonperturbative~\cite{RSSV00} estimates), the gap 
increases from a few to several tens of MeV over a small range of
temperature around $T=80$~MeV. 
The CFL rates are compared to free quark-antiquark annihilation  
at order $\alpha_S^0$ in a baryon-rich QGP 
first computed in ref.~\cite{CFR87}:
\begin{equation}
\frac{dR_{q\bar{q}\to ee}}{d^4q}=
\frac{\alpha^2}{4\pi^4}\frac{T}{q} \ f^B(q_0;T)
\sum\limits_{q=u,d,s} e_q^2 \
{\mbox {ln}}\frac{(x_{-}+{\mbox {exp}}[-(q_0+\mu_q)/T])(x_{+}+
{\mbox {exp}}[-\mu_q/T])}{(x_{+}+{\mbox {exp}}[-(q_0+\mu_q)/T])(x_{-}+
{\mbox {exp}}[-\mu_q/T])} \ ,
\label{dRdM2_0}
\end{equation}
with $\alpha=e^2/4\pi$ and 
$x_{\pm}={\mbox {exp}}[-(q_0\pm q)/2T]$.\footnote{One should note that 
perturbative corrections arising from new processes allowed in a QGP
at finite baryon density can give substantial 
enhancement over the lowest-order (${\cal O}(\alpha_S^0)$) result of
eq.~(\ref{dRdM2_0}) in the low-mass region ($M<1$~GeV) region~\cite{MG01}.
Qualitatively, such corrections have a similar shape to the ones
induced by finite gaps.} 
As characteristic for soft emission rates generated from in-medium effects,  
the gap-induced processes exhibit a much steeper slope than the perturbative 
$q\bar{q}$ annihilation, exceeding the latter towards small masses. 
It is straightforward to incorporate the leading (free) antiparticle 
contribution within the CFL one-loop results by retaining the antiparticle 
projectors in the diagonal parts of the Nambu-Gorkov propagator.
\vskip 0.4cm 
\begin{figure}[ht]
\vskip 0.1cm 
\bce
\epsfig{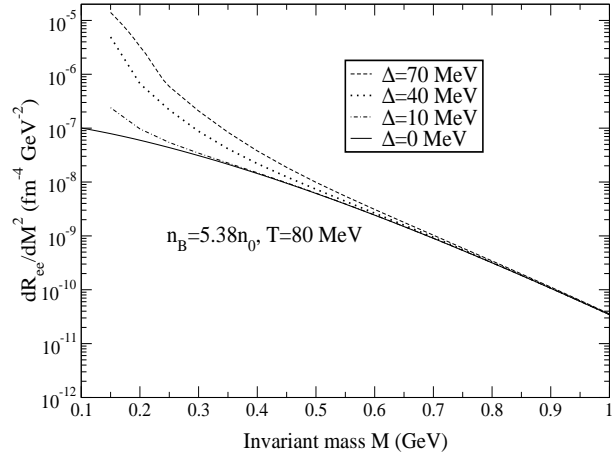}
\ece
\caption{Dilepton emission rate including the antiparticle
contribution.(For $\Delta=0$ MeV, the perturbative $q\bar{q}$
annihilation rate is recovered.)}
\label{figpslep}
\end{figure} 
\noindent
Upon doing so we recover the limit of perturbative $q\bar{q}$ annihilation 
as the gap goes to zero, cf.~Fig.\ref{figpslep}, where also 
the additional processes due to finite pairing gaps have been
included.  
Whereas the gap acts as a threshold energy
($q_0 \geq 2\Delta$) in the response function, eq.(\ref{response}), 
this is not so for the invariant-mass spectrum, as there is no constraint 
on the three momentum transfer $q$.  Nevertheless, the energy threshold 
leaves its traces also in the invariant mass spectra. To illustrate this
point, we have magnified in Fig.~\ref{figps4} the region of very low
masses $M\le 0.2$~GeV.   
\noindent An inflection point of the rate is discernible at $M\simeq 2\Delta$, below
which a finite 3-momentum transfer is required, entailing  
a suppression of dilepton production. In fact, the rate with the smaller
pairing gap even exceeds the one with the more tightly bound Cooper pairs.

\begin{figure}[!bh]
\vspace{0.8cm}
\bce
\epsfig{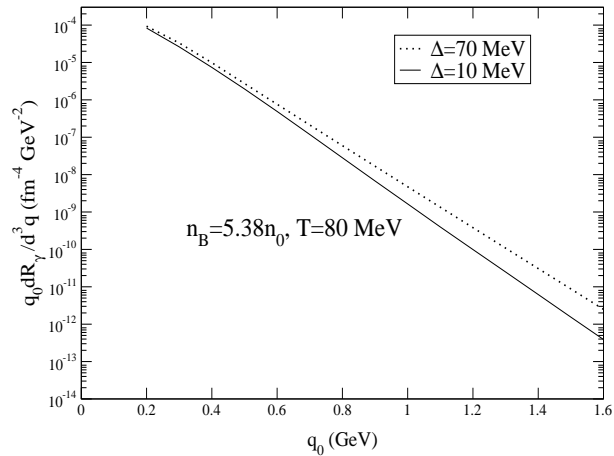}
\ece
\vspace{0.3cm}
\caption{Direct photon emission rates from gapped quarks in the CFL phase at pairing
gaps of 10 and 70~MeV (full and dotted lines, respectively).}
\label{figps6}
\end{figure}

The photon emission rate, eq.~(\ref{Rphot}), from the CFL
phase at the same density and temperature ($\mu_q=350$~MeV, $T=80$~MeV) 
is plotted in Fig.~\ref{figps6}, for different choices of the gap value.In the forthcoming sections, we will compare these rates to that from a hot
hadron gas with in-medium effects included.

\section{CFL Phase and E.M. Current Correlator II: Hidden Local Symmetry }
\label{sec_cc2}
\subsection{Generalized Mesons and Effective Lagrangian}
Beyond the 1-loop result from the previous section, 
the existence of mesonic bound states in the vector and axial vector channel
in high density QCD~\cite{RSWZ00} is expected to further affect the dilepton
emissivity from dense low temperature matter.
The knowledge of the coupling of these modes to the electromagnetic current 
affords us the opportunity to complement studies of dilepton 
emission from hot and dense hadronic environments with ``hadronic-type'' 
rates from the high density side. 
Starting from a suggestion made in ref.~\cite{RSWZ00}, we 
will here adopt the Hidden Local Symmetry (HLS) approach to construct  
the coupling of these so-called  'generalized' vector mesons in the CFL phase 
to the photon field. As an effective low energy theory, the HLS framework 
determines the couplings between generalized vector and pseudoscalar 
mesons, the latter being the Goldstone modes of chiral symmetry breaking
in the CFL phase. This feature also provides the answer to the question 
whether dilepton production through the generalized vector mesons bears any
overlap with the quark one-loop description of the previous section.
It turns out that the quark-based (weak coupling) analysis, carried out
to leading logarithm accuracy, does not support a coupling to 
the generalized pseudoscalar states (though it may be finite at
next-to-leading order) and moreover, the coupling of the vector
excitation to the physical vector current is zero~\cite{RSWZ00}. 
Therefore, the decay of the 
generalized vector mesons (or, equivalently, the annihilation of 
generalized pseudoscalar mesons) adds to the dilepton rate independent
of the one-loop result of the previous section. The HLS framework allows
us to determine the coupling of the generalized vector mesons to the
electromagnetic current, and hence the contribution to the dilepton rate.
\vskip 0.1cm
The idea of interpreting low-lying vector (and axial vector) mesons in free
space as dynamical gauge bosons of a hidden local symmetry originated in
refs.~\cite{Bando,Kugo,Uehara,Yamawaki}. 
 
A particular advantage of this framework becomes apparent in the 
conventional $[SU(2)_L\otimes SU(2)_R]_{\rm global}$ nonlinear
sigma model where the electromagnetic interaction can be implemented as
a global gauge symmetry, coupling to global isospin charge, distinct
from the (hidden) local $SU(2)_V$ charge to which the rho meson
($\rho$) couples. This separation of the photon source from the $\rho$
allows for a natural explanation of several phenomenological facts (such as
$\rho$ dominance of the vector coupling to pions, KSRF relations and
universality of the $\rho$ coupling) at the cost of fixing one parameter
(the universal vector coupling constant).

\vskip 0.1cm 
In weak coupling, the (composite) generalized vector mesons in the 
CFL phase are distinct from the screened and Higgsed gluons~\cite{RSWZ00}. 
Could they be the realization of a hidden local symmetry, besides
the explicit local color symmetry? If the Cooper pairs have finite size,
there is, strictly speaking, no hidden local invariance in the order
parameter apart from explicit color symmetry. However, we will work in
the zero size approximation, in which case the condensate does indeed
exhibit a hidden local symmetry corresponding to the unbroken color-flavor 
SU(3) group\footnote{A similar caveat holds for effective field
descriptions of hadrons in the nonperturbative vacuum; it may be overcome 
by higher order terms in the chiral perturbation series.}. Just like the 
$\rho$ meson in case of hidden local symmetry at zero density~\cite{Uehara},
the vector mesons of the CFL phase are composite and heavy.
In this respect, our hidden gauge approach is different from
ref.~\cite{Casal} wherein the local $SU(3)_c$ was the hidden (color) gauge.  
In the following we construct an effective Lagrangian for the
Nambu-Goldstone bosons arising from the spontaneous symmetry breaking,
with the vector mesons introduced as gauge particles of the unbroken local
symmetry. As in the vacuum case, they are rendered physical degrees
of freedom by postulating a kinetic term (assumed to be generated at the
quantum level), which does not affect the low energy dynamics
of the theory. 

\vskip 0.1cm 
To start the derivation of the HLS Lagrangian in the CFL ground state,
we recall that, formally, a nonlinear sigma model based on the coset 
manifold $G/H$ is gauge
equivalent to another model with $G_{\rm global}\otimes H_{\rm local}$
symmetry and that the gauge bosons corresponding to the hidden local
symmetry in $H_{\rm local}$ are composite fields. For the CFL phase,
$G\equiv [SU(3)_C\otimes SU(3)_L\otimes SU(3)_R]_{\rm global}$ and
$H\equiv [SU(3)_{C+L+R}]_{\rm local}$. In terms of (right-handed) quark 
fields, the explicit form of the order parameter reads
\begin{equation}
\langle q_{R,\alpha}^{ai} q_{R,\beta}^{bj} \rangle 
=\frac{1}{2} (C^\dag \gamma_5 P_R)_{\alpha\beta} (\kappa_1 \delta^{ai}
\delta^{bj} + \kappa_2 \delta^{aj} \delta^{bi} ) \quad,
\label{order}
\end{equation}
where $\{a,b\}$, $\{i,j\}$ and $\{\alpha,\beta\}$ denote color, flavor and
spinor-indices, respectively ($C$: charge conjugation matrix,  $P_R$:
right-handed chiral projector). An equivalent form applies to the left-handed 
fields. Following refs.~\cite{Ho99,RSWZ00}, this symmetry-breaking pattern
implies the emergence of 8 Goldstone bosons corresponding to  
$SU(3)_{c+A}$-valued condensate rotations of the form 
${\rm e}^{-i\gamma_5\Theta^A T^A}$ with associated group generators 
$T^A$ ($A=1,\dots,8$). In what follows we will assume the CFL ground state
to be invariant under the parity transformation. We introduce unitary fields 
corresponding to the individual left- and right-handed rotations, 
characterized by expectation values   
\begin{equation}
\langle \Phi_{Lai}\rangle = -\langle \Phi_{Rai}\rangle =
\delta_{ai} \ ,  
\label{parity}
\end{equation}
which implies $\kappa_1=-\kappa_2$ in eq.~(\ref{order}), and fixes the 
overall $U(1)$ phases in 
\begin{equation}
\Phi_{L,R} \equiv {\rm e}^{2i\phi_{L,R}} \ U_{L,R}(x)    
\end{equation}
at, e.g., $\langle\phi_L\rangle=0$ and $\langle\phi_R\rangle=\pi/2$ 
\footnote{We will not include in our low energy description the breaking 
of $U(1)_B \rightarrow Z_2$ associated with baryon superfluidity ($H$-meson) or
the anomalous $U_A(1)$ current ($\eta^{\prime}$ meson), as they are not
relevant to the problem at hand.}. The generalized left- and right-handed 
$SU(3)$ fields can be parameterized by 
\begin{eqnarray}
U_{L}(x) &=& {\rm e}^{2i\pi_L(x)/F_{\pi}} \\
U_{R}(x) &=& {\rm e}^{2i\pi_R(x)/F_{\pi}} \ , 
\end{eqnarray}
where $\pi_{L,R}=\pi_{L,R}^A T^A$, and the $SU(3)$ generators are
normalized as ${\rm Tr}(T^A T^B) = \delta^{AB}/2$. 
At this point we have doubled the number of colored (generalized) Goldstone 
bosons represented by $SU(3)$ phases of $U_L$ and $U_R$.
The latter transform under $G$ as $U_L\rightarrow g_c^* U_L g_L^{\dag}$ and
$U_R\rightarrow g_c^* U_R g_R^{\dag}$, respectively. 
The following further decomposition, 
\begin{eqnarray}
U_{L}(x) &=& \xi_{c}^{T}(x)\xi_{L}(x)
\\
U_{R}(x) &=& \xi_{c}^{T}(x)\xi_{R}(x) \ , 
\end{eqnarray}
then makes explicit the additional invariance under the
hidden local symmetry $h\in [SU(3)_{C+L+R}]_{\rm local}$:
\begin{eqnarray}
\xi_{c}^{T}(x)&\rightarrow & \xi_{c}^{T}(x)h^{\dag}(x) \\
\xi_{L,R}(x)&\rightarrow & h(x)\xi_{L,R}(x) .  
\end{eqnarray}
We will fix the
local $SU(3)_{C+L+R}$ gauge by choosing $\xi_c^T(x)=\xi_{L,R}(x)\equiv \xi(x)$.
It follows that $\pi_L=\pi_R$, so that we recover 
the expected number of Goldstone bosons (eight). This simply means
that color locks to the flavor diagonal so that left- and right-handed
'pions' are the same. In this way, we implement the description of
(generalized) pions as phases of the (colored) diquark condensate, 
and avoid an explicit appearance of the gluon fields in our effective
theory\footnote{In other works, it was the field $\Sigma=U_L^{\dag} U_R$ 
which mapped the meson octet, with gluons acquiring a mass via
the gauge fixing of color as the hidden local symmetry. This
integrating out of gluons led to an identification with the effective
chiral Lagrangian at zero density.}.  With this realization, one 
introduces a vector gauge field $V_{\mu}$ that transforms
inhomogeneously under the hidden local symmetry parametrized by
$h(x)$. Since the gauge fields may be viewed as connections in the
coset manifold, we can define the gauge covariant derivative as
\begin{equation}
D_{\mu}\xi(x) = \partial_{\mu}\xi(x) -iV_{\mu}\xi(x) \ .
\end{equation}
The effective Lagrangian for the CFL pions is now assembled  
from invariants under $SU(3)_C\otimes SU(3)_L \otimes SU(3)_R
\otimes[SU(3)_{C+L+R}]_{\rm local} \otimes (\mbox{parity})$:
\begin{eqnarray}
{\cal L}_{eff} = &-&\frac{F_{T}^2}{4}{\rm
Tr}(D_{0}\xi^{\dag}\cdot \xi - D_{0}\xi\cdot \xi^{\dag})^2 -
\frac{F_{S}^2}{4}{\rm Tr}(D_{i}\xi^{\dag}\cdot \xi - D_{i}\xi\cdot
\xi^{\dag})^2  \label{Loz} \\  
&-&a_{T}\frac{F_{T}^2}{4}{\rm Tr}(D_{0}\xi^{\dag}\cdot \xi +
D_{0}\xi\cdot \xi^{\dag} -2iV_{0})^2
-a_{S}\frac{F_{S}^2}{4}{\rm Tr}(D_{i}\xi^{\dag}\cdot \xi +
D_{i}\xi\cdot \xi^{\dag} -2iV_{i})^2 + \dots \nonumber , 
\end{eqnarray}
where the breaking of Lorentz invariance of the finite density ground state
is reflected by individual temporal and spatial constants 
$F_T$ and $F_S$, respectively\footnote{One has
$F_S/F_T=v_{\pi}=1/\sqrt{3}$ where $v_{\pi}$ is the velocity of the 
Goldstone modes~\cite{Son}. 
To leading logarithm accuracy, the pion decay constant 
$F_T\simeq \mu_q/\pi$~\cite{RSWZ00,Son,Zar,Mir}.}. 
The first two terms of eq.~(\ref{Loz}) reproduce the kinetic term for the
pions while the other two (proportional to $a_T$ and $a_S$) vanish 
identically utilizing the equation of motion for the $V_{\mu}$ field,
\begin{equation}
V_{\mu}^a = -i{\rm Tr}\{T^a (\partial_{\mu}\xi \cdot
\xi^{\dag}+\partial_{\mu}\xi^{\dag}\cdot \xi)\} \quad.
\end{equation}
As in the zero density case, the vector field propagates
by postulating a kinetic term assumed to arise from interactions on 
the microscopic QCD level. This does not affect the limit of low energy 
dynamics that the effective Lagrangian is aiming to describe. The ellipsis 
in eq.~(\ref{Loz}) denotes higher order and possible mass terms of heavier
excitations consistent with the CFL symmetries, which we ignore. 
With the addition of the gauge kinetic term, and expanding up to
${\cal O}(p^2)$ ($p$: momentum of the pion field), we have
\begin{eqnarray}
{\cal L}_{eff} &=&\frac{1}{2}(\partial_{0}\boldpi)^2
+\frac{v_{\pi}^2}{2}(\partial_{i}\boldpi)^2
+\frac{a_T}{2} \boldV_{\!\!0} \cdot \boldpi \times \partial_{0} \boldpi
+\frac{a_S}{2} \boldV_{\!\!i} \cdot \boldpi \times \partial_{i} \boldpi 
\nonumber\\
&& +\frac{1}{2}(a_T F_{T}^2) \boldV_{\!\!0}^2 + 
\frac{1}{2}(a_S F_{S}^2) \boldV_{\!\!i}^2
+\frac{1}{2g_T^2} (\boldF_{\!\!0i})^2
-\frac{1}{2g_S^2} (\boldF_{\!\!ij})^2 + \dots \ ,  
\label{Leff2}
\end{eqnarray}
where we define the standard field strength tensor 
$\boldF_{\!\!\mu\nu}=\partial_{\mu} \boldV_{\!\!\nu}-\partial_{\nu}  
\boldV_{\!\!\mu} + \boldV_{\!\!\mu} \times \boldV_{\!\!\nu}$\footnote{Here, 
boldface notation indicates 8-component vector in flavor-space with 
the cross- ($\times$) and dot-products to be read as 
$(\boldV \times \boldV)_A = f_{ABC} V^B V^C$ and  
$\boldV \cdot \boldV = V^A V^A$ with $f_{ABC}$ the antisymmetric 
structure constant of the $SU(3)$ group.}.
Upon rescaling the vector fields, $\boldV_{0}\rightarrow g_T \boldV_{0}$, 
$\boldV_i\rightarrow g_S \boldV_i$, we identify 
\begin{eqnarray}
g_{V_0\pi\pi} &=& \frac{1}{2} \ a_T \ g_T \ , \quad 
m_{T}^2 = a_T \ g_T^2 \ F_{T}^2 \ ,
\label{ksfrT}
\\
g_{V_i\pi\pi} &=& \frac{1}{2} \ a_S\  g_S \ , \quad 
m_{S}^2 = a_S \ g_S^2 \ F_{S}^2 \ .
\label{ksfrS}
\end{eqnarray}
We note that, apart from Lorentz symmetry breaking, the effective Lagrangian, 
eq.~(\ref{Leff2}), is formally identical to that obtained at zero density with 
hidden local symmetry for the chiral $SU(2)_{V}$ subgroup.
For the numerical calculations presented below, the value of $g_T$ will be
taken to be that of $g_{\rho\pi\pi}$ in vacuum\footnote{Another estimate 
of $g_T$ may be obtained by a study of the 4-pion vertex by integrating out 
the $\rho$ fields in the infinite mass limit.}. We will show later 4-dimensional 
transversality of the photon self energy requires $g_S=g_T$.  Both the mass of 
the vector field and its coupling to two generalized pions depends on the 
parameters $a_T$ and $a_S$. A suggestive choice will emerge when we introduce
electromagnetism into our effective theory. This is our next step, 
so as to assess the coupling of the generalized (neutral) vector mesons 
to the (modified) photons with application to e.m. spectra.

The true electromagnetic gauge boson in the CFL medium is
the modified photon $(\tilde{A}_{\mu})$, with a modified coupling
$\tilde{e}$ to charged degrees of freedom.  $\tilde{A}_{\mu}$ is the
gauge field associated with local $U(1)_{\tilde{Q}}$ phase transformations 
on the $U_{L,R}$ fields. Noting that this is equivalent to a simultaneous 
transformation
$\xi_f\rightarrow \xi_f e^{-ieQ{\rm cos}\theta}$, $\xi_c\rightarrow \xi_c
e^{-igY{\rm sin}\theta}$, we see that $U_{L,R}$ transform as
\begin{equation}
U_{L,R}\rightarrow {\rm e}^{i\tilde{e}Q} U_{L,R} {\rm e}^{-i\tilde{e}Q} \ ,
\end{equation}
where we have used $Q=-Y$ for multiplication from the left. Thus, the
covariant gauging with electromagnetism can be achieved via  
\begin{eqnarray}
 D_{\mu}^{em}\xi_f &=& \partial_{\mu}\xi_f +i\tilde{e}\xi_f \tilde{A}Q
 \\ D_{\mu}^{em}\xi_c^{*} &=& \partial_{\mu}\xi_c^{*}
 +i\tilde{e}\xi_c^{*} \tilde{A}Q \ 
\end{eqnarray}
(the superscript $*$ denotes complex conjugation). Thus the effective 
Lagrangian, eq.~(\ref{Leff2}), with electromagnetism incorporated becomes  
\begin{eqnarray}
{\cal L}_{eff}^{em} = {\cal L}_{eff} 
&-& \frac{1}{4} (F_{\mu\nu}^{{em}})^2  - \frac{2\tilde{e}m_T^2}{\sqrt{3}g_T}
V_{0}^{8}\tilde{A}_{0} 
- \frac{2\tilde{e}m_S^2}{\sqrt{3}g_S}V_{i}^{8}\tilde{A}_{i}
+ \frac{1}{2}\frac{4\tilde{e}^2m_T^2}{3g_T^2}\tilde{A_0}^2
+ \frac{1}{2}\frac{4\tilde{e}^2m_S^2}{3g_S^2}\tilde{A_i}^2 
\nonumber \\
&-&(\frac{a_T-2}{\sqrt{3}})\tilde{e}\tilde{A_{0}}f_{AB8}\pi^A\partial_{0}\pi^B
-(\frac{a_S-2}{\sqrt{3}})\tilde{e}\tilde{A_{i}}f_{AB8}\pi^A\partial_{i}\pi^B
+ \dots  \ .  
\label{Lgauge}
\end{eqnarray}
It obviously exhibits generalized $\rho^0$-$\tilde{\gamma}$ mixing ($V^8$ is 
assigned the generalized neutral $\rho$ meson).
Vector meson dominance (VMD), i.e., the absence of a direct $\pi\pi\gamma$
vertex, is recovered for $a_T=a_S=2$, in analogy  
to free space, where it successfully reproduces the empirical $\rho$ 
dominance of the electromagnetic pion form factor. 
Such a parameter choice thus seems natural on account of the formal 
resemblance of the CFL and vacuum case at the level of the bare 
effective Lagrangian. On the other hand,  it has recently been 
argued~\cite{harada}, based on a one-loop renormalization
group equation (RGE) analysis, that VMD (for $N_f$=3) is only accidentally 
realized in the chirally broken phase at zero density, and that it 
does not hold towards the phase boundary of chiral restoration. 
The CFL phase, however, is a priori of a rather different nature 
(i.e., its symmetries and physical excitations), so that VMD cannot
be precluded by the perturbative arguments invoked in ref.~\cite{harada}. 

With VMD enforced, the mass eigenstates of the photon and the vector meson 
fields follow from eq.~(\ref{Lgauge}) via diagonalization,
\begin{eqnarray}
B_{\mu}&=&(g^2+\frac{4}{3}\tilde{e}^2)^{-1/2}
\biggl(g\tilde{A}_{\mu}+\frac{2}{\sqrt{3}}\tilde{e}V_{\mu}^{8}\biggr)
\\
W_{\mu}^{8}&=&(g^2+\frac{4}{3}\tilde{e}^2)^{-1/2}
\biggl(gV_{\mu}^{8}-\frac{2}{\sqrt{3}}\tilde{e}\tilde{A}_{\mu}^{8}\biggr)
\\
m_{B}^{2}&=&0 \ , \quad m_{W^8}^{2}=aF^{2}(g^2+\frac{4}{3}\tilde{e}^2) \ ,
\end{eqnarray}
where $g=g_T,F=F_T$ for $\mu=0$ and $g=g_S,F=F_S$ for $\mu=i$. 
We see that $\rho$-$\tilde{\gamma}$ mixing spontaneously breaks
$[SU(3)]_{hidden} \otimes U(1)_{Q}$ down to $U(1)_{em}$. 

We have thus shown that the HLS framework, which has proven 
successful in the chirally broken vacuum state,   
can be implemented into the (chirally broken, $N_f$=3) high density color 
superconducting ground state as well. The (generalized) $\rho$ meson 
emerges as a resonance of generalized two-pion states, with its mass and 
width determined by the effective Lagrangian eq.~(\ref{Lgauge}). 
We are now in position to evaluate its contribution to the
dilepton emission rate in dense CFL matter.

\subsection{Dilepton and Photon Rates}
Adopting the VMD hypothesis for the coupling of (real and virtual) CFL 
photons to the generalized isovector mesonic current (as motivated in the 
previous section), the electromagnetic correlation function can be saturated 
by the $\rho^0$ meson. The pertinent dilepton rate, 
eq.~(\ref{Rdilep}), then takes the form  
\begin{equation}
\frac{dR_{ee}}{d^4q}=-\frac{\alpha}{3\pi^3 M^2} \
\tilde{\alpha}\, f^{\rho}(q_0;T)\biggl[\frac{(m_{T})^4}{g_T^2} 
g_{00} \ {\rm Im} D^{00}(M,q) + \frac{(m_{S})^4}{g_S^2} 
g_{ij} \ {\rm Im} D^{ij}(M,q)\biggr] \ 
\label{VDMrate}
\end{equation}
with $\tilde{\alpha}=\tilde{e}^2/4\pi$, $f^{\rho}$ the thermal 
Bose-Einstein occupation factor, and 
$D^{\mu\nu}$ the 
in-medium propagator of the generalized $\rho$ meson, which is summed over its
longitudinal and transverse polarizations to obtain the rate.  
The form of this propagator follows from inverting the equations of
motion for the spatial and temporal components of the field $V_{\mu}$,
\begin{equation}
D^{\mu\nu}=\frac{P_T^{\mu\nu}}{M^2-m_S^2}
+\frac{P_L^{\mu\nu}}{M^2-m_S^2\frac{q_0^2}{M^2}+m_T^2\frac{q^2}{M^2}}
+\frac{q^{\mu}q^{\nu}}{m_T^2q_0^2-m_S^2q^2} \ .
\label{rhopropmf}
\end{equation}
Here, $P_L$ and $P_T$ are the standard longitudinal and transverse
projectors for photons in matter~\cite{lebellac}.
We note that for $m_T=m_S=m_\rho$, this propagator reduces to that for 
the massive spin-1 field in the covariant (Landau) gauge, 
\begin{equation}
D^{\mu\nu}=\frac{1}{M^2-m_\rho^2}\biggl(-g^{\mu\nu}
+\frac{q^{\mu}q^{\nu}}{m_\rho^2}\biggr) \ .
\end{equation}
Though we have made a particular gauge fixing, the gauge parameter
would appear only in the $q^{\mu}q^{\nu}$ term of
eq.~(\ref{rhopropmf}), and therefore not contribute
to the rate by transversality of the photon. The rate is a gauge invariant 
(observable) quantity, driven by the imaginary parts of the longitudinal 
and transverse modes, i.e., spectral functions. The transverse propagator
 has a pole at $M=m_S$, while the longitudinal propagator admits two
 positive definite solutions. These are given by
\begin{equation}
m^2_{\pm}=\frac{m_S^2}{2}\biggl[1\pm \sqrt{1-4q^2\frac{m_T^2-m_S^2}{m_S^4}}\biggr]\quad.
\end{equation}
As $m_T > m_S$ (cf.~footnote 6), there exists a limiting 3-momentum
given by $q_{\rm max}=m_S^2/(2\sqrt{m_T^2-m_S^2})$
beyond which the longitudinal mode is damped. The lower mode corresponding
to the choice of negative sign in the above equation vanishes at
$q=0$ where it has zero strength, and only the upper mode persists,
becoming degenerate with the transverse part, as it must in  
the rest frame. 

\begin{figure}[ht]
\vskip 1cm
\bce
\epsfig{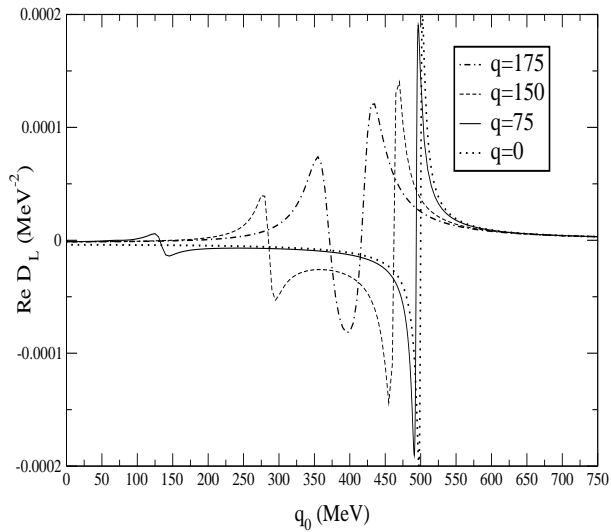}
\ece
\vskip 0.3cm
\caption{The real part of the longitudinal $\rho$ propagator as a
function of frequency for different 3-momenta $q$ at fixed masses
$m_S$ and $m_T$. The upper and lower modes 
are degenerate at $q=q_{\max}$ and separate with decreasing $q$.}
\label{rerhops} 
\end{figure}

\begin{figure}[ht]
\vskip 1cm
\bce
\epsfig{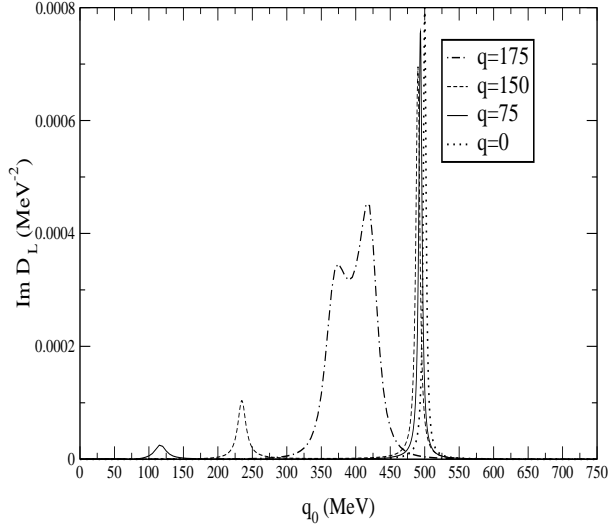}
\ece
\vskip 0.3cm
\caption{The imaginary part of the longitudinal $\rho$ propagator for 
fixed $m_S$, $m_T$ and various values for the 3-momentum $q$.
The upper ('+')  mode is dominant except when $q$ is very close to 
$q_{\rm max}=176.5$ MeV.}
\label{imrhops} 
\end{figure}

We may view the longitudinal mode as having an energy-momentum dependent 
self-energy by decomposing the inverse propagator as  
\begin{eqnarray}
M^2-m_S^2\frac{q_0^2}{M^2}+m_T^2\frac{q^2}{M^2} &\equiv&
M^2-m_S^2-\Sigma_L(M,q)
\\
\Sigma_L(M,q)&=&-(m_T^2-m_S^2)\frac{q^2}{M^2} .
\end{eqnarray}
We impose retarded boundary conditions by continuing the $1/M^2$ factor 
as $q_0\rightarrow q_0-i\epsilon$. 
To illustrate the above features, we insert a small (finite) imaginary part 
into the self energy and display real and imaginary part of the longitudinal 
propagator in Figs.~\ref{rerhops} and \ref{imrhops}, respectively, for fixed 
values of the masses ($m_S=500$ and $m_T=\sqrt{3}m_S$). 
\vskip 0.3cm
For $q$ approaching $q_{\rm max}$ ($\simeq 176.5$ MeV with our choice of masses), the
two modes  merge while for smaller $q$ they are distinct. For
$q=0$, only the mode degenerate with the transverse part survives. 
The retarded boundary condition renders the spectral function, 
\begin{equation}
{\cal A}_L(\omega)=-2\,{\rm Im} D_L(\omega)\quad,
\end{equation}
a positive definite quantity (which directly figures into the e.m. rates), 
even though the lower ('-') mode has a negative discontinuity. 
We have checked that the spectral sum rule for the two
longitudinal modes,
\begin{equation}
\int_{-\infty}^{\infty}\frac{d\omega}{2\pi} \omega{\cal A}_L(\omega)=1
\end{equation}
holds for independent values of $m_T$ and  $m_S$.

\begin{figure}[!h]
\vskip 1cm
\bce
\epsfig{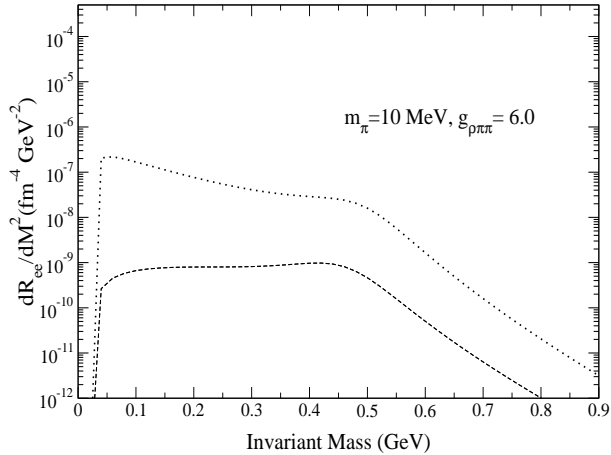}
\ece
\vskip 0.3cm
\caption{Invariant-mass spectrum from the longitudinal (dashed) and transverse (dotted) modes of propagation of the $\rho$ meson in the CFL phase (pion mass $m_{\pi}= 10$ MeV) 
for $\mu_q=350$ MeV, $T=80$ MeV and $g_{\rho\pi\pi}$ set to the vacuum value
of 6.0. The broad peaks of the longitudinal and
transverse mode are distinct and discernible around 475~MeV.}
\label{rhops} 
\end{figure}

\begin{figure}
\vskip 1cm

\bce
\epsfig{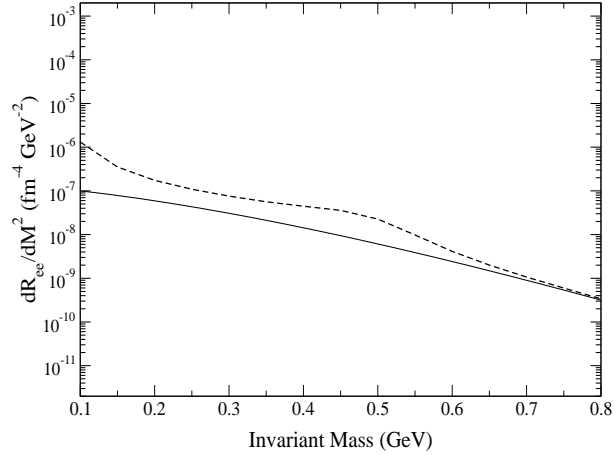}
\ece
\caption{Dilepton rate (dashed line) including the peak from the
generalized $\rho$ meson for $\mu_q=350$ MeV, $T=80$ MeV and
$\Delta=10$ MeV. The solid line represents the free $q\bar{q}$
annihilation rate at order $\alpha_S^0$ at that temperature and
density. The $\rho$ peak enhances significantly the rate from
the paired quarks.}
\label{totalps}
\end{figure}

We can go beyond the mean field results, eq.~(\ref{rhopropmf}), by including
loop corrections from the coupling of the $\rho$ to 2 pions. 
In that case, the imaginary part of the $\rho$ self-energy diagram 
(decay width) are ascribed to the $\rho\pi\pi$ interaction, which we
evaluate according to the $\rho\pi\pi$ matrix element determined by 
eq.~(\ref{Leff2}).  The self energy is generated to one-loop order 
from the squared matrix element integrated over
available phase space and summed over final states,
\begin{equation}
-{\rm Im} \Sigma_{\pi\pi}(q_0,q)
=\frac{1}{2} \sum\limits_f \int d\Pi_f\,({\cal TT^{*}})  \ , 
\label{self} 
\end{equation}
with the reduced matrix element 
${\cal T}=ig_{\rho\pi\pi}\epsilon_{\rho}^{\mu}(2q+k)_{\mu}$, 
$\epsilon_{\rho}^{\mu}$
the $\rho$ polarization vector and  $g_{\rho\pi\pi}=g_{T,S}$ for $\mu=0,i$,
respectively. Focusing on the decay of the $V^8$ component 
($\rho^0$), which alone couples to the photon, the only nonzero antisymmetric 
structure constants are $f^{128}$ and $f^{458}$, 
representing two independent decay channels which must be
added incoherently. The imaginary part of the components of the self energy 
as computed from eq.~(\ref{self}) are then 
\begin{eqnarray}
{\rm Im}\Sigma_{\pi\pi}^{00}(M) &=& {\rm Im}\Sigma_{\pi\pi}^L(M)(P_L^{00}+P_T^{00})
\nonumber\\ 
{\rm Im}\Sigma_{\pi\pi}^{0i}(M) &=& \sqrt{{\rm Im}\Sigma_{\pi\pi}^L(M){\rm
Im}\Sigma_{\pi\pi}^T(M)}(P_L^{0i}+P_T^{0i})
\nonumber\\
{\rm Im}\Sigma_{\pi\pi}^{ij}(M) &=& {\rm Im}\Sigma_{\pi\pi}^T(M)
 (P_L^{ij}+P_T^{ij}) \quad, 
\end{eqnarray} 
with
\begin{eqnarray}
{\rm Im}\Sigma_{\pi\pi}^L(M) &=& -\frac{g_T^2~M^2}{32\pi}
\nonumber\\ 
{\rm Im}\Sigma_{\pi\pi}^T(M) &=& -\frac{g_S^2~M^2}{32\pi} \ .
\end{eqnarray} 
We now see that the self-energy is not transverse
($q_{\mu}\Sigma^{\mu\nu}\neq 0$) unless $g_S=g_T$. This observation
enforces the equality of the two couplings (denoted henceforth by $g$), 
and therefore the equality of the self-energies (henceforth denoted by 
$\Sigma_{\pi\pi}$).  The self-energies are resummed through a 
Schwinger-Dyson equation,
yielding longitudinal and transverse $\rho$-propagators of the form  
\begin{eqnarray}
D_L(M, q) &=&
\frac{1}{M^2-m_S^2\frac{q_0^2}{M^2}+m_T^2\frac{q^2}{M^2}-
{\rm Re}\Sigma_{\pi\pi}-i{\rm Im}\Sigma_{\pi\pi}} \ ,
\label{DL2} \\
D_T(M,q) &=& \frac{1}{M^2 - m_S^2 - {\rm Re}\Sigma_{\pi\pi}
-i{\rm Im}\Sigma_{\pi\pi}} \ .
\label{DT2}
\end{eqnarray}
The real parts of the self energy are more involved and could be
obtained from subtracted dispersion relations. We will assume that
they do not significantly alter the pole structure. 
Combining eqs.~(\ref{VDMrate}), (\ref{DL2}) and 
(\ref{DT2}) provides an expression for the thermal emission rate (as a
function of the four-momentum $q^{\mu}$) from the
$\rho$ contribution. 
Numerical integration over 3-momentum yields the rate $dR/dM^2$ as a function
of invariant mass, cf.~Fig.~\ref{rhops}, where we also included a finite pion mass. 
The pion mass arises from a non vanishing current 
quark mass matrix and has been estimated at around $m_\pi\simeq10$~MeV 
to leading log accuracy within a Bethe-Salpeter approach for the pionic 
bound state energy~\cite{RSWZ00}. 
It simply introduces an extra phase space factor of $(1-4m_{\pi}^2/M^2)^{3/2}$
into the imaginary part of the self-energy, which leads to a notable
suppression of the rate only for invariant masses $M\lsim 50$~MeV. 
Beyond that, the $M^2$ dependence of the (imaginary part of the) 
self energy causes a slight downward shift of the transverse $\rho$ peak below 
its nominal mass at $M=m_S$ (this feature might not persist upon inclusion
of corrections to the real part of $\Sigma_{T}$). A close inspection
of the denominator in the longitudinal propagator reveals that it also 
peaks somewhat below $M=m_S$ in the invariant-mass spectrum. For very
large $M (>1~{\rm GeV})$, the ratio of the rates (transverse to
longitudinal) approaches a constant value of two, as expected.
Numerical values for $m_S$ and $m_T$ follow from eqs.~(\ref{ksfrT})
and ~(\ref{ksfrS}) with VMD enforced ($a_T=a_S=2$), and $F_T$ given to
leading logarithm accuracy by~\cite{RSWZ00}
\begin{equation}
F_T^2\simeq\frac{\mu^2}{\pi^2}\frac{8x_0^2+\pi^2(1-{\rm exp}(-2x_0))}{4x_0^2+\pi^2} \ 
 ,  \qquad x_0=\frac{3\pi^2}{\sqrt{2}g} \ .
\end{equation} 
The magnitude of the generalized $\rho$-meson contribution is compared to 
the perturbative quark rate in Fig.~\ref{totalps}. 
The former dominates over the latter in the invariant-mass region up to 
about 0.7~GeV, emphasizing the role of composite (i.e., generalized mesonic) 
degrees of freedom. Even the one-loop CFL corrections to the quark rate as 
calculated in sect.~\ref{sec_cc1} can compete with effective theory results  
only for sufficiently large pairing gaps, cf.~Figs.~\ref{figps3} and \ref{figps4}.

\section{COMPARISON TO LOW-DENSITY APPROACHES}
\label{sec_comp}
The evaluation of the electromagnetic emissivities in the previous
two sections applies to densities large enough to support the formation
of superconducting quark matter. More specifically,
we assumed the Cooper pairing to follow the symmetry pattern of color-flavor
locking. Towards lower chemical potentials the color superconducting matter
will undergo a (possibly first order) transition into ordinary hadronic 
matter\footnote{For realistic values of the strange quark mass,
$m_s\simeq 100-150$~MeV, the 
CFL phase might well turn into the 2-flavor superconductor (2SC) before
hadronic matter is recovered; the one-loop photon self energy for the 2SC 
state could be calculated along similar lines as done in sect.~\ref{sec_cc1}, 
whereas the HLS construction of Sect.~\ref{sec_cc2} does not apply due to the
absence of chiral symmetry breaking. On the other hand, if the 2SC window on 
the $\mu_q$ axis does not exist, the possibility for a continuous transition
from CFL to hadronic matter has been raised. Such a scenario would in fact
corroborate the evidence for duality that we are about to discuss in terms
of the electromagnetic response.}, which, nevertheless, is likely to occur 
at densities well above that of normal nuclear matter where hadronic medium 
effects will be substantial. A natural question to ask then is, how the CFL 
emissivities compare to results from dense hadronic phases at low 
temperatures, which should be thought of as an extrapolation from the 
low density side. 
\begin{figure}[!bh]
\vskip 1cm
\bce
\epsfig{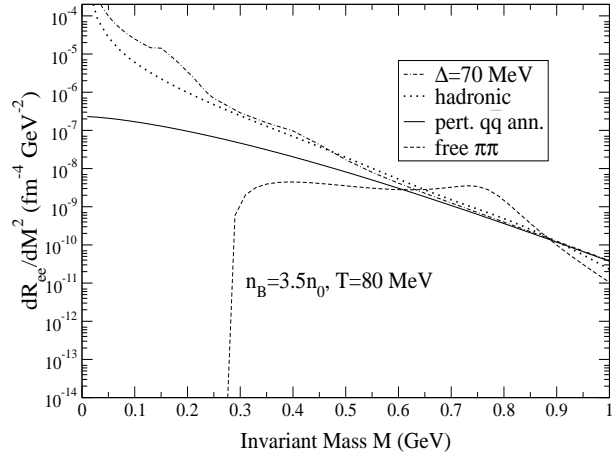}
\ece
\caption{Comparison of dilepton rates in the in medium hadronic and high density 
approaches, suggestive of a quark-hadron duality picture holding in dense matter.}
\label{quarkhadps}
\end{figure}
Similar comparisons have been discussed before for both 
photon and dilepton rates at large temperatures ($T=150-200$~MeV) 
with~\cite{RW99,Ra01} and 
without~\cite{KLS91,LR98,LYZ98} the inclusion of baryon density effects.  
The low temperature and high density dilepton case (for $T=80$~MeV and 
$n_B=3.5n_0$ corresponding to a quark chemical potential of $\mu_q=290$~MeV) 
is exemplified in Fig.~\ref{quarkhadps}, where the CFL rates 
(the combined contribution from the previous two sections) are  
confronted with in-medium hadronic calculations based on 
ref.~\cite{RW99,UBRW00},
as well as the pertinent lowest order results, i.e., perturbative 
${\cal O}(\alpha_s^0)$ $q\bar{q}$ as well as free $\pi^+\pi^-$
annihilation. 
The latter two are obviously very different in shape, 
but upon the inclusion of interactions, it appears intriguing that the 
rates from the two in principle very different ground states exhibit a 
remarkable tendency to coincide within a factor of two (somewhat
more towards very low masses below 300~MeV or so).
This agreement opens the possibility that the dilepton signal from the 
baryon-dense hadronic phase smoothly matches the description within the plasma
phase, i.e., the `bottom-up' and 'top-down' extrapolations
support the emergence of duality. 

Another important issue in this context is the one of chiral symmetry 
restoration. This obviously requires an evaluation of the axial vector
current correlation function. Although chiral symmetry in the CFL 
phase is broken, it has been found in ref.~\cite{RSWZ00} that, to leading 
logarithm accuracy, the masses of the generalized vector and axial vector
mesons are degenerate, $M_V=M_A\simeq 2\Delta (1-{\rm e}^{-C/g})^{1/2}$ 
with  $C=(3-\sqrt{3})\pi^2\sqrt{6}$.
This is possible due to the more involved 
symmetry breaking pattern (as compared to the usual vacuum case), 
which, e.g., does not require a finite value for the two-quark condensate,
but only a nonzero chiral four-quark condensate.  The leading log 
$\rho$-$a_1$ degeneracy is also consistent with our earlier 
statement that to this accuracy, there is no coupling of the vector
current to two-pion states. Although the simplest version of the HLS 
approach (as constructed in Sect~III) does not incorporate explicit 
axial vector degrees of freedom, it can be accordingly extended~\cite{BFY88},
which will not be attempted here.
We furthermore note that model 
calculations~\cite{ARW99,RSSV00} indicate certain chiral breaking 
effects in the CFL ground state to be substantially weaker than in the 
vacuum. For e.g., 'constituent' quark masses turn out to be about an order of 
magnitude smaller than the pairing gaps $\Delta\simeq 100$~MeV. 
On the other hand, the pion decay constant, 
which also constitutes an order parameter of spontaneous chiral breaking, 
is parametrically large, ($F_T^2=3~F_S^2\simeq \mu_q^2/\pi^2$),
i.e., of order ${\cal O}(g^0 \mu_q^2)$. Thus the question arises how 
this can be reconciled with the Weinberg sum rules~\cite{Wei67}. 
One of them relates the pion decay constant to the integrated difference 
in spectral strengths for vector and axial vector correlators, 
\begin{equation}
f_\pi^2= -\int \frac{dq_0^2}{\pi} \ \frac{1}{q_0^2} \ 
\left( {\rm Im}  \Pi_V(q_0)  - {\rm Im}  \Pi_A(q_0) \right)    
\label{wsr1}
\end{equation}
and has been shown to hold in the medium as well~\cite{KS94}.
A near degeneracy of Im~$\Pi_V$ and Im~$\Pi_A$ implies a small pion 
decay constant. This discrepancy has been resolved in ref.~\cite{MT01}, 
being associated with genuine two-loop diagrams involving an 
intermediate gluon propagator ('dumbbell diagram').   
Naively, these diagrams are of higher order in $g$, but an infrared 
singularity in the gluon propagator, being regulated upon resummation 
by the standard Debye-/Meissner masses, turns them into order 
${\cal O}(g^0)$ contributions. This lifts the $V$-$A$ degeneracy 
and restores the sum rule eq.~(\ref{wsr1}).  

\begin{figure}
\vskip 1cm
\bce
\epsfig{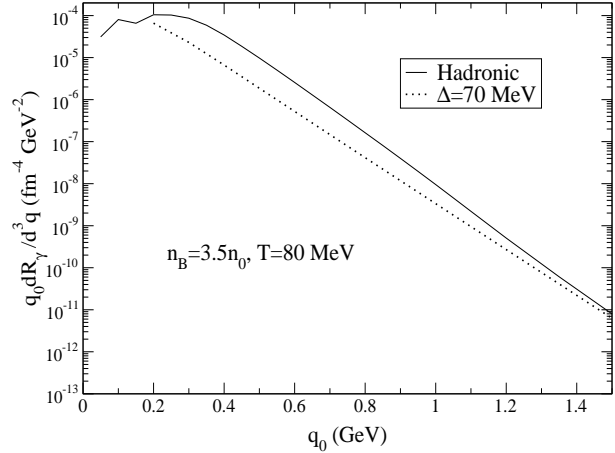}
\ece
\vskip 0.3cm
\caption{Photon rates from CFL quark matter and a dense hadronic 
environment at equivalent densities.}
\label{photonps} 
\end{figure}

Finally, turning to thermal photon production, Fig.~\ref{photonps} confronts
rates from the CFL phase with in-medium hadronic results~\cite{RW99}.
Here, too, the emission pattern from the dense quark system
is quite similar to that computed in hadronic or QGP scenarios. 

\section{Conclusions}
\label{sec_concl}
In this article we have investigated the electromagnetic response 
of color superconducting quark matter for three massless flavors 
in the so-called CFL realization. Two approaches were pursued  
to calculate the e.m. current correlator. 

The first one consisted of a perturbative one-loop calculation in terms 
of the appropriate quark (hole) quasiparticles. In contrast to the 
free Fermi gas case, where imaginary parts arising from Landau damping
are purely spacelike, the presence of finite quark gaps
induces a non-vanishing strength (imaginary part) in the timelike 
region, thus opening up the possibility for dilepton production.
More quantitatively, taking as a baseline the free quark-antiquark 
annihilation process, quark gaps on the order of 50~MeV lead to 
appreciable contributions in the 3-momentum integrated dilepton rate
up to invariant masses as large as 400~MeV.     

The second approach was based on effective field methods utilizing the
notion of spontaneous chiral symmetry breaking in the CFL ground state.
After identifying the suitable set of eight generalized Goldstone bosons, 
a chiral Lagrangian has been constructed with massive vector mesons 
introduced using the hidden local symmetry technique. Vector meson dominance 
arises as a natural parameter choice within this framework. The resulting  
dilepton rates, corresponding to the annihilation of generalized pions 
via generalized $\rho$ mesons, receive contributions from both the
spin-transverse and -longitudinal modes, the former being dominant at all
invariant masses. Both modes have a typical mass
around 500~MeV.

Furthermore, we argued that the absence of a pion coupling to the vector 
current in the leading logarithm approximation implies that both perturbative
and effective Lagrangian contribution add to the e.m. correlator.
Extrapolating the obtained dilepton rates down to moderate densities,
we find them to be comparable (within a factor of 2 or so) with 
in-medium hadronic calculations as available in the literature.
The two generic features shared in both pictures are 
(i) a strong low mass enhancement originating from soft many-body 
excitations, and (ii) a rather 
structureless 'continuum' beyond invariant masses of about 0.5~GeV. 
This extends the notion of 'quark-hadron duality' into the baryon-dense 
regime at low temperatures, similar to what has been suggested before
for the high temperature case. Our analysis can and should be extended
to the 2SC phase, as well as the crystalline phases of dense QCD at low
temperature~\cite{OVER,LOFF}. The results should prove useful for a way to 
probe and discriminate the features of 
the QCD phase diagram at low temperature and moderate
densities in a way that complements detailed hadronic calculations.
 Optimistically, one may even hope to use electromagnetic
signals in future heavy-ion experiments for a potential access to the
high density regime at moderate temperatures.

\vskip 0.5cm

\section*{ACKNOWLEDGEMENTS}
We thank Thomas Sch\"afer for 
discussions and M. Harada for a useful comment. This work was supported by the US-DOE grant DE-FG0288ER40388.


\begin{flushleft}

\end{flushleft}
\end{document}